\PassOptionsToPackage{svgnames,dvipsnames}{xcolor}
\documentclass[runningheads,a4paper]{llncs}

\usepackage[square,sort,comma,numbers]{natbib}
\usepackage{xcolor}
\usepackage[hidelinks]{hyperref}

\hypersetup{
  pdftitle={Emergent Behaviour in Financial Markets},
  pdfauthor={Omar Inverso, Emilio Tuosto, Dragisa Zunic}
}

\newcommand\comment[1]{}

\renewcommand{\textcolor}[2]{#2}

\author{
  Omar Inverso \and
  Emilio Tuosto \and
  Dragi{\v{s}}a {\v{Z}}uni{\'c}
}

\institute{
  Gran Sasso Science Institute, Italy
}

\begin{document}

\title{Emergent Behaviour in Financial Markets\thanks{This paper has been accepted at the International Symposium on Leveraging Applications of Formal Methods, Verification and Validation (ISoLA 2026), in the Resourceful Engineering of Complex and Autonomous Systems (ReoCAS) track.}}%:\\{\color{purple}A Computational Perspective} }

\maketitle

\begin{abstract}
Some properties of so-called complex or collective systems
can be observed to emerge from the interactions of elementary agents.
This phenomenon, known as emergent behaviour, has long since been studied 
in the most diverse disciplines, with
recent growing awareness from the formal methods community
about the opportunity of opening up to seemingly distant disciplines
with appropriate technology for computer-aided reasoning.
Different peculiar elements of complexity make automated reasoning on these systems
particularly challenging.
We consider electronic financial markets to drive our
discussion.
We identify and structure the sources of complexity to tackle in order
to provide computational support for the analysis of emergent
phenomena.
We refrain from evaluating the suitability of specific
technical solutions or frameworks of preference, which would as usual
require simplifying assumptions and divert from the actual phenomenon
of interest.  
Rather, we elaborate on possible alternatives to handle some of the
main technical aspects involved in automated analysis, while retaining
a solid and concrete interpretation of the domain,
\textcolor{blue}{and in doing so outline a more systematic research
  program for the formal specification and analysis of market
  mechanisms.}
\end{abstract}

%!TEX root = 2026-ISoLA-Economics.tex

%--------- --------- --------- --------- --------- --------- --------- --------- --------- --------- 
\section{Introduction}
%--------- --------- --------- --------- --------- --------- --------- --------- --------- --------- 
Many natural or artificial systems may be seen as large conglomerates of \emph{agents} that
repeatedly perform elementary \emph{actions} in a shared \emph{environment} based on
limited knowledge about each other.
Despite the simplicity of the agents,
surprisingly sophisticated collective dynamics can emerge in the system.
This phenomenon, known as \emph{emergent behaviour}, has long since been fascinating 
scholars from the most diverse disciplines.
For instance,
some species of foraging ants in a colony are able to gradually discover
the shortest path while transporting food from a location back to their nest,
thereby solving a mathematical optimisation problem~\cite{goss_self-organized_1989}.
Emergent phenomena can be observed in ethology~\cite{norris1988cooperative},
social networks~\cite{DBLP:conf/socialcom/HaglichRP10,DBLP:journals/jocss/SasaharaCPCFM21},
%economic systems~\cite{RePEc:aea:jecper:v:3:y:1989:i:4:p:85-97,DBLP:journals/alife/Tesfatsion02},
artificial intelligence \cite{doi:10.1073/pnas.79.8.2554}, and more generally
in several classes of so-called complex or collective systems
\cite{ferscha2015collective,Mataric:1993:DEB:171174.171225,grimm2005,Bonabeau1999}.

%--------- --------- --------- --------- --------- --------- --------- --------- --------- --------- 
Improving our understanding of such systems can yield significant impact.
Among the other things, it is worth to mention the studies on consensus and opinion polarisation
in social groups via agent-based models~\cite{DBLP:journals/jasss/HegselmannK02},
the remarkable number of applications and theoretical results in optimisation inspired
by the above-mentioned collective intelligence of ants~\cite{DBLP:journals/cim/DorigoBS06}, and
swarm-inspired engineering techniques adopted for instance
in the design of space systems at NASA to increase resilience~\cite{DBLP:journals/itpro/VassevSRH12}
and for synthetic cell-like microparticles for biological applications~\cite{doi:10.1073/pnas.2426850122}.

The study of complex systems poses several challenges, such as
predicting emergent properties given the behaviour of the agents,
inferring the behaviour of the agents by observing the system as a whole,
assessing convergence, stability, and many other properties.
At a theoretical level,
all this is usually quite difficult due to inherent factors of complexity,
most notably the large or unbounded size of the system in terms of the number of agents,
the fact that at any given time the system can have multiple possible configurations, and
the well-known intricacies introduced by concurrency.
At a practical level, the difficulty is further exacerbated by technological shortcomings,
despite the considerable advancements over the last few decades 
in terms of modern parallel hardware, computing infrastructures, and
general-purpose analysis techniques  and decision procedures.
In particular, there seems to be a considerable shortage of computer-aided instruments for rigorous reasoning,
especially in the form of efficient, scalable, and easy to use software tools \emph{for domain experts}.
%This prevents opening up towards other disciplines where there might be a great deal to be gained.

%\tocheck{
%It has been argued that the main hindrances to successful application of automated workflow
%for the analysis of complex systems are
%the linguistic barrier,
%i.e. the fact that a domain expert may find it difficult to specify the system under study
%with generic languages 
%(1) the necessity of domain-specific languages to describe the system of interest naturally and intuitively 
%(i.e. without syntactic tricks) and
%(2) the technically-challenging re-use of general-purpose artefacts.
%}

The opportunity for the relevant areas of research in computer science
to open up to other disciplines has already been argued,
in e.g.~\cite{DBLP:journals/tosem/StefanoNI22}.
Preliminary work has been carried out towards automated analysis of emergent properties
for different classes of systems,
e.g. stability of matchings between groups of peers~\cite{DBLP:conf/coordination/NicolaDIT17},
collective behaviour in multi-robot systems~\cite{DBLP:journals/firai/NicolaSI18,DBLP:conf/isola/InversoTT20},
group cohesion and related properties in flocking birds~\cite{DBLP:conf/isola/NicolaSIV22},
convergence to efficient path in foraging ants~\cite{DBLP:conf/cmsb/NicolaSIV23}, and
synchronisation in applauding audiences~\cite{DBLP:conf/isola/StefanoI24}.
These contributions consider notable instances of emergent behaviour
from a variety of systems,
demonstrating possible workflows for automated analysis, but
by doing so inevitably introduce assumptions and simplifications with respect to the system and property of interest.

In this paper, we explore the intricacies of modelling and analysing
complex systems through the lens of electronic financial markets.
We argue that a successful approach to the specification and analysis
of complex systems should be grounded on solid interdisciplinary
efforts, the reuse of state-of-the-art general-purpose techniques and
tools, and a clear balancing between the separation of concerns among
the different phases of the specification and the analysis on the one
hand, and the adaptations required in each phase to improve efficiency
on the other hand.
% Added by dragisa, as response to reviewer
\textcolor{blue}{The perspective adopted in this paper is not limited to explaining why emergent behaviour in electronic markets is difficult to analyse. It also points toward a more concrete methodological direction: the development of formal and executable accounts of market mechanisms, together with reusable workflows for specification, verification, and simulation. In this sense, the discussion is not only about clarifying the problem space, but also about outlining a more systematic research programme.}

\paragraph{Structure of the paper.}
Sect.~\ref{sec:efm} surveys electronic financial markets and some
of their properties focusing on those related to emergent behaviour.
Sect.~\ref{sec:discussion} elaborates on the features, challenges, \textcolor{blue}{and methodological implications} posed by complex systems to modeling and verification.
Sect.~\ref{sec:conc} reports our concluding remarks.

%%% Local Variables:
%%% mode: LaTeX
%%% TeX-master: "2026-ISoLA-Economics"
%%% End:

%!TEX root = 2026-ISoLA-Economics.tex
%--------- --------- --------- --------- --------- --------- --------- --------- 
\section{Electronic Financial Markets}\label{sec:efm}
%--------- --------- --------- --------- --------- --------- --------- --------- 

Electronic financial markets are among the core systems of modern
finance. They are computerised trading mechanisms that guide the
interaction between supply and demand, namely they process orders (e.g., buy and sell orders) submitted by agents.
The role of the market mechanism is to determine
- according to well-established economic principles -
how these orders are prioritized, matched, and translated into
transactions at a certain price. Different market designs implement
this coordination process through different priority, timing, and
matching~rules~\cite{ohara95,Milgrom2021}.
% \eMcomm[check]{
%   % Two of the most important paradigms in use in modern
%   % finance are the Continuous Limit Order Book (CLOB) and Frequent Batch
%   % Auctions (FBA).
%   %
%   % In the CLOB model, 
%   The most widespread paradigm in use in modern finance is CLOB, after
%   Continuous Limit Order Book.
%   %
% }
% %
% \eMcomm[new]{ A key characteristic of CLOB is that time is treated as
%   continuous; before discussing the implications in terms of emergent
%   behaviour (cf. \cref{sec:eeb}), we briefly summarise the historical
%   evolution which led to CLOB (cf. \cref{sec:his}).
% }\marginpar{\tiny Dragisa: there is no mention of FBA here; it should be.}

% \eMcomm[rm]{
% Arguably, neither of these models fully captures the nature of the computation taking
% place on a trading venue, i.e., the computational structure of interaction
% underlying market activity. Most notably, the computational core guiding this
% interaction may need to incorporate elements of concurrency and parallelism
% already at the level of fundamental design. This connects to a long-standing
% market-design challenge: how to structure trading so that speed, fairness,
% liquidity, and price discovery are balanced in a robust way ~\cite{Biais2005,Budish2015,Aquilina2022}.
% }

\subsection{
  % Evolution: market models through a computational lens
From trading floors to electronic markets}\label{sec:his}
The evolution of trading venues can be viewed as a gradual evolution
of how markets organise the interaction among agents, i.e., market participants.
Traditional exchanges relied on \emph{trading floors}, where traders
interacted physically through voice, gesture, and local social
dynamics.
Later, electronic markets replaced this environment with automated
matching engines, most notably using the continuous limit order book
(CLOB) model.
According to this model, the market evolves sequentially and
continuously: orders are processed one-by-one as soon as they arrive
in the market following a price-time priority.
The state of the market is therefore updated incrementally after each
individual interaction.

For decades this model worked reasonably well. However, with the rise
of high-frequency trading, speed itself became a major strategic
advantage. Faster participants could exploit tiny timing differences,
race to react to new information, and systematically extract value
through latency arbitrage and the sniping of stale liquidity quotes. 
\textit{Latency arbitrage} refers to profit opportunities created by tiny differences in reaction time, when fast market participants are able to act before the majority of participants, i.e., before prices across the market adjust. \textit{Sniping} refers to the rapid execution of a trade against an order before the trader who posted that quote has time to update or cancel it after new information arrives.

These phenomena became widely known through the story of IEX
and Lewis's \emph{Flash Boys}~\cite{Lewis2014}, and were later studied
formally.
Notably, Budish et al. introduced the frequent batch auctions model
(FBA)~\cite{Budish2015}.
Unlike CLOB, FBA assumes discrete time so as to collect orders
arriving within a short time interval and process them
simultaneously as a batch.
Therefore, FBA periodically processes a joint clearing outcome for all
orders in the batch at once~\cite{Budish2015}, rather than matching
continuously order-by-order.
This relates to one of the fundamental principles in financial
economics, the so-called \emph{efficient market hypothesis}, according to which prices should
reflect available public information~\cite{Fama1970}. Latency
arbitrage in continuous markets exposes a misalignment with this ideal: even public
information may be monetized through tiny speed advantages before prices fully
adjust. In this sense, batch-based mechanisms can be interpreted as an attempt
to reduce the role of millisecond reaction time and shift competition from speed
toward~price. Regulatory attention is also present in this matter~\cite{SEC2010MarketStructure}.

Recent systems such as OneChronos continue this broader shift toward
auction-based and computationally richer trading mechanisms~\cite{Milgrom2021}.
Similar concerns also arise in decentralized finance, where issues
such as front-running, maximal extractable value (MEV), and
transaction-ordering manipulation motivate alternative coordination
and ordering mechanisms in order to achieve
fairness~\cite{Daian2020FlashBoys,FerreiraParkes2023,Bartoletti2025}.

% environment
In most modern electronic markets, trading agents are aware of the common observable \textit{environment}: current bids and asks, visible liquidity, trade history, public information, and the rules of the trading mechanism. By submitting, canceling, or modifying orders, they also change this environment for the next round of interaction. This feedback between market state, agent behaviour, and mechanism rules gives rise to different classes of market properties.

% PROP
\subsection{Properties of financial markets}\label{sec:eeb}
In the context of markets, we consider properties of different
nature. Some are core mechanism properties and they follow directly
from the design. Such properties may be suitable for formal
specification and
verification~\cite{CervesatoKhanReisZunic2019}. Other properties
concern strategic agents' behaviour enabled by the design, i.e., rules
of the game. For example, alignment~\cite{itz26}, latency arbitrage, the possibility of sniping, etc.
These
properties depend on how agents exploit or react to the mechanism and
are therefore harder to treat as simple design invariants. A further
class consists of emergent properties of the market as a whole. Some
of them concern market quality, such as liquidity depth, bid--ask
spreads, execution quality, or stability. Others are empirical
regularities learned from data, such as the stylised facts of
financial markets\footnote{A stylised fact is a simplified,
  big-picture pattern that is widely accepted as true because it holds
  up across different times and markets when abstracting away the
  outliers.}.
In either case, these
properties do not follow solely from the local rules of the market
model. They also depend on agents' behaviour and interaction, or
appear as empirical regularities observed in market data.

\paragraph{Core properties} are those that follow directly from the
design of the market model. At the same time, they express minimal
requirements for the system. Main core properties are that (i) the
market is never left in a crossed state, meaning that after the model
has processed the relevant orders, the bid (most competitive/maximal
buy) is strictly lower than the ask (most competitive/minimal sell)
and that (ii) trade always takes place between the orders with highest
priority at that given moment.
The latter property also implicitly guarantees that order priority is
respected at all times, and that the actual trade price is directly
related to the current bid or ask price.

\paragraph{Strategic properties} are induced by market design and emerge as strategies that
agents apply given the market. In CLOB, the sequential processing of orders,
coupled with treating time as continuous, creates potential opportunities for value extraction based on speed~\cite{Budish2015,Aquilina2022}. Examples of strategic properties are: (i) reduction of front-running against regular investors; (ii) reduction of sniping, understood as predatory execution against stale liquidity-provider
quotes. As said, these properties are not simple invariants of the market core: they
depend on how agents observe the market, react to information, and compete
under the timing and priority rules of the mechanism.
% For this reason, they are
% typically analysed using economic, game-theoretic, or empirical arguments,
% rather than by formal verification alone.

\paragraph{Market-quality properties} concern the market as a whole, and arise indirectly from both the market design and the behaviour of agents given the market design. Notable examples are: (i) improved execution prices for regular investors; (ii) improved conditions for liquidity providers; furthermore, as a consequence of these, we have 
(iii) increased liquidity and narrower bid--ask spreads; and arguably (iv) improved
stability. These properties arise indirectly
from the rules of the game and from the way agents exploit, adapt to, or are
constrained by those rules. For instance, a mechanism that reduces sniping or
latency arbitrage may change or improve the incentives of liquidity providers, which can
increase market depth, decrease spreads, and improve execution quality.
% These properties usually require economic, empirical, or simulation-based analysis.

\paragraph{Observed empirical properties} are recurring regularities
uncovered by analysing historical data across
markets~\cite{cont2001empirical}. These properties yield well-known
examples of emergent behavior in economic systems. In particular, let
us mention: (i) absence of simple predictability, meaning that past
returns have little power to predict future price direction, so prices
are difficult to forecast locally even though aggregate regularities
may still be observed; (ii) ``heavy tails'', meaning that large
price movements occur more often than one would expect, so crashes and
spikes are not merely exceptional accidents but recurrent features of
market dynamics; and (iii) the volume--volatility relation, meaning
that high trading activity is often associated with high volatility,
so prices tend to move more when many agents trade or react at the
same time.

\paragraph{Square-root law of price impact} is a peculiar example of
an observed property. This law states that executing a buy or sell
meta-order of total size $Q$ moves the price on average by an amount
approximately proportional to $\sqrt{Q}$~\cite{Toth2011}.  It gives a
simple and explicit quantitative relation between traded volume and
price movement. In simple terms, larger orders have larger impact, but
the impact grows sublinearly rather than linearly. Remarkably, the
average impact appears to depend mainly on the total traded volume,
and only weakly on the precise way in which the order is split or
executed over time~\cite{Bucci2019,Maitrier2025}.

%%% Local Variables:
%%% mode: LaTeX
%%% TeX-master: "2026-ISoLA-Economics"
%%% End:

%!TEX root = 2026-ISoLA-Economics.tex

%--------- --------- --------- --------- --------- --------- --------- ---------
\section{Discussion}\label{sec:discussion}
%--------- --------- --------- --------- --------- --------- --------- ---------
Trading venues are not the only cornerstone financial system, in the realm of conventional finance, that has
attracted interest from theoretical computer science and formal methods. A
well-known example is the work of Peyton Jones et al. on financial contracts,
where precise and compositional descriptions of complex derivative contracts
are given~\cite{Jones2000}. Trading venues provide a complementary challenge. While financial contracts
highlight the need for compositional semantics of complex obligations over
time, markets require reasoning about open-ended interaction among many agents
under fixed rules. This makes trading venues a useful case study for identifying the sources of
complexity that arise when applying formal reasoning to economic systems with
strategic interaction and emergent phenomena.

\textcolor{blue}{The point of the present discussion goes beyond enumerating the sources of complexity: it is also to indicate how they bear on the possibility of a more disciplined methodology for formal market-core engineering. In other words, the very challenges identified here are at the same time the ingredients from which a more concrete research program must be built.}

From a computational standpoint, an electronic financial market (Sect.~\ref{sec:efm})
can be viewed as a concurrent system,
where economic agents are standalone processes with their own behaviour.
For instance, FBA addresses speed races in CLOB by processing orders
in discrete batches rather than sequentially as they arrive, which at the same time relates to the efficient-market ideal.
From a computational viewpoint, this is no surprise:
races naturally arise in asynchronous concurrent systems.
As discussed in practical terms later on in this section,
concurrency itself is already a source of trouble for automated analysis.
In addition, other specific characteristics that make electronic financial markets 
particularly interesting systems can represent further elements of complexity.
In the rest of this section, we briefly overview some of them, and
discuss how they affect different aspects of formal analysis in practice.

\subsection{Features}
Much like other so-called collective or complex systems,
different classes of \emph{emerging properties} of potential interest
are not directly related to the behaviour of the agents,
but may indirectly emerge from that.
An interesting related aspect of global system dynamics is \emph{nonlinearity},
i.e. small local changes can lead to large changes in the system as a whole.

Another distinctive feature is \emph{nondeterminism}.
Some systems may have variable initial conditions at the level of the environment
(e.g. liquidity, etc.) that in the actual model need to be accounted for.
Nondeterminism may also be necessary to model the behaviour of the agents,
for example to model internal choices such as the possibility of picking one in a set of
multiple available actions (e.g.  bid or sell).

Electronic financial markets may exhibit \emph{non-trivial interaction} mechanisms,
e.g. many-to-many matchings among dynamically formed groups.
Interaction (or communication) may be further complicated 
by the fact that the identity of the agents composing the system may not be relevant or known, or that agent identities are abstracted away,
a feature that is commonly referred to as~\emph{anonymity}.

Economic systems like financial markets
are considered \emph{large} given their \emph{size}
in terms of number of agents composing them, and \emph{open},
because the size itself is not fixed
as new agents can join or leave the system at any time;
this feature is occasionally referred to as \emph{open-endedness}. 
A financial market is also usually \emph{heterogeneous}
given the presence of different kinds of agents in it,
e.g. investors with different behavioural strategies, competing actors, etc.
On the other hand, no agent can have special privileges or can control or coordinate the others,
meaning that there is \emph{no~leader}.

\subsection{Specifications}
We have already argued that software tools for \emph{domain experts}
can represent a great opportunity to open up towards other disciplines. 
Obviously, any workflow for automated analysis would require the object of study
(in this case, the system and properties of interest) to be described in a machine-tractable format.
To a domain expert, such specification phase will represent the first barrier
to successful adoption of any integrated workflow for formal specification and analysis of systems.

Early approaches based on \emph{top-down specification}
try to describe the system as a whole in terms of aggregate features.
Different mathematical frameworks have been relied upon to study biological systems,
among others. For instance, flocking birds have been modelled using graph
theory~\cite{DBLP:journals/tac/Olfati-Saber06},
distributed control laws~\cite{DBLP:journals/ijcon/ShiWC09},
statistical mechanics~\cite{bialek2012statistical}, and differential equations~\cite{doi:10.1073/pnas.0711437105}.
Ants colonies have been studied by means of process algebras~\cite{sumpter_ants_2001},
search algorithms and strategies~\cite{MONMARCHE2000937,traniello_foraging_1989}, and
with probabilistic techniques~\cite{deneubourg1983probabilistic}.
Such approaches may not be very intuitive and
are often out of reach for domain experts with limited experience in mathematical formalisms. 

Compositional methodologies for \emph{bottom-up specification}
focus on formalising the individual components rather than the entire system have gained prominence.
The system is defined in terms of the agents composing it along with their individual rules;
the collective behaviour of interest is not explicitly formalised and
may or may not emerge from an interleaved sequence of such rules.
This has been done in a variety of disciplines,
including epidemiology~\cite{DBLP:conf/iccS/KuylenLBH20},
ecology~\cite{DBLP:journals/siamma/FinkelshteinKK09},
economics~\cite{DBLP:journals/alife/Tesfatsion02,stiglitz2011}, and social sciences \cite{cedermanPNAS2002}, and
seems to fit more naturally the compositional nature of complex systems.

As a separate parallel trend in computer science,
an abundance of expressive formalisms
for specifying system and properties of concurrent systems
have been developed over the last few decades
in the area of formal methods and languages.
Modular verification of safety properties of concurrent systems via theorem proving
has been proposed in~\cite{DBLP:conf/pldi/PadonMPSS16,DBLP:journals/pacmpl/SergeyWT18}.
Well-engineered frameworks~\cite{DBLP:books/aw/Lamport2002} have been used
for verifying linear temporal properties via explicit-state model checking~\cite{DBLP:conf/charme/YuML99},
symbolic model checking~\cite{DBLP:journals/pacmpl/0001KT19}, and theorem proving~\cite{DBLP:conf/fm/CousineauDLMRV12}.
These formalisms aim to maximise the potential areas of applications
of state-of-the-art technology (in some cases developed with them), and
for this reason are general purpose.
Despite the expressive power,
a general-purpose language will not necessarily be sufficiently intuitive for domain experts,
discouraging them as potential end-users.

More recent efforts from the formal methods area at large
try to provide potential end-users with domain-specific languages
with explicit constructs to take into account
the peculiar characteristics of a given class of systems.
For instance, interaction in electronic financial markets
does not rely on the identity of agents.
Traditional point-to-point communication primitives like send-receive
implemented in many modern programming languages for distributed systems would thus be
unsuited to capture naturally this kind of interaction.
It has already been shown how tailored communication primitives
can express more naturally sophisticated interaction mechanisms in collective systems,
such as stigmergy in ants~\cite{DBLP:conf/cmsb/NicolaSIV23} and 
attribute-based communication among dynamically formed groups of peers in social media~\cite{DBLP:journals/scp/NicolaDIT18};
%which would otherwise be tricky to model in terms standard communication operations
%such as point to point, broadcast, or multicast.
on the other hand, under the hood this requires plenty of ingenuity and tradeoffs
for the actual language implementation to be effective in practice~\cite{DBLP:journals/scp/NicolaDIT18}.
In general,
the development of a domain-specific language
would require non-negligible interdisciplinary effort 
in order to identify appropriate linguistic elements and clearly define their semantics.

\subsection{Scalability}
The primary concern in the automated analysis of a concurrent system
is usually the remarkable size of the state space.
Process interleaving is one of the main sources of complexity in that respect.
Due to concurrency, the individual actions of economic agents can intertwine, 
causing the number of possible system evolutions to grow very quickly with
the number of agents and actions for each agent.
Let us consider an elementary system where multiple agents
operate according to plain sequences of actions,
each action consisting in buying or selling a commodity.
With only three agents and four actions for each agent,
there would be about $3.4 \cdot 10^4$ feasible execution traces to cover
for exhaustive analysis.
With just one more agent and one more action per agent, 
the number of feasible execution traces
would grow to about $11.7 \cdot 10^9$. %11 732 745 024
In practice,
the size of a financial market is realistically expected
to be larger than the examples above by orders of magnitude.
In addition, 
as financial markets are typically open systems,
new agents can dynamically join in, which
implies unbounded parallelism that exacerbates the problem.

In addition to process interleaving, other sources of nondeterminism
add further complexity.
Nondeterminism can also be introduced to account for internal choices of agents
(e.g. a choice between buying or selling a commodity), to express
multiple conditions (e.g. initial conditions for an agent or at the level of the environment), and
to model communication asynchrony (e.g. message-passing systems where
messages between agents can be delivered out of order).
A single binary choice (e.g. to model the tossing of a coin)
can cause the overall state space to grow twice as big in the worst case;
a non-deterministic variable of arbitrary bitwidth 
would contribute an extra exponential factor to the overall state space size. 

The unbounded time horizon for property checking is an orthogonal issue.
Some of the properties introduced in Sect.~\ref{sec:efm} (e.g. market stability)
may in principle require unbounded system observations to be proven,
which is usually unfeasible in practice.
For instance, experiments e.g. on formal verification of simplified models of ant colonies
have shown that the analysis can get quickly unfeasible after a few \emph{epocs},
i.e. steps where each agent can take an action.

These shortcomings may be addressed in part by leveraging
existing mature general-purpose technology for \emph{symbolic} analysis,
where the actual reasoning on the system is carried out in terms of sets 
of possible system behaviours rather than enumeratively.
With respect to the size and time horizon unboundedness,
it may be worthwhile exploring possible adaptations of existing techniques
such as property-directed reachability~\cite{DBLP:conf/vmcai/Bradley11},
or novel techniques inspired by size cutoffs
in the style of parameterised model checking~\cite{Kouvaros2015} and
completeness thresholds for under-approximate analysis~\cite{DBLP:journals/iandc/KonnovVW17}.
Practical approaches include complementing 
the results of exhaustive analysis on a system of small size and time horizon
with simulation on a larger scale~\cite{DBLP:conf/isola/NicolaSIV22}.

\subsection{Integration}
To date,
simulation is still the most widely used approach to reason about classes of systems with collective behaviour.
Mature simulation platforms are available e.g.
for multi-agent systems~\cite{Tisue04netlogo:a,DBLP:journals/simulation/LukeCPSB05} and
multi-robot systems~\cite{Koenig2004,PinciroliARGoSmodularparallel2012,Rohmer2013}. 

As argued, simulation is only marginally effective as it typically
suffers from limited coverage due to process interleaving and
nondeterminism.  Approaches to simulate or formally verify very
specific properties on population protocols of unbounded size have
been proposed in~\cite{DBLP:conf/cav/BlondinEJ18}; it is worth to
remark that agents in population protocols are all of the same kind
and necessarily very simple.  In contrast, many complex systems
--including financial markets-- have expressive properties and agents of
heterogeneous and arbitrary kind.  Explicit-state model checking has
been proposed for multi-agent
systems~\cite{DBLP:conf/kbse/BordiniDFF08,DBLP:journals/sttt/LomuscioQR17}.
It is well-known that explicit-state model checking is prone to state
space explosion in the presence of non-deterministic variables.
%Moreover, the current implementations are tied to specific languages and back ends for the actual reasoning.
Agent-based economic systems have been studied via statistical model checking
too~\cite{DBLP:conf/datamod/VandinGLC21,BlandoFagioloGiachiniVandinIvanaj2026,VANDIN2022104458}.
A common limitation of these approaches is their rigidity; frameworks
that are flexible both in terms of specification languages and
analysis techniques would be more desirable.

The specification language is no longer a strongly limiting factor in semantics-based verification,
because the semantics of the specification language becomes part of the specifications.
The K framework integrates semantics-based language specification, execution, and
formal analysis~\cite{DBLP:conf/oopsla/StefanescuPYLR16}.
%In principle, one can define the syntax and operational semantics of any language, and
%automatically get a reachability checker for programs written in that language.
%However, K requires rewrite-based rules for the operational semantics; adapting existing semantics may entail non-negligible effort.
%K relies on symbolic execution for the analysis, with ongoing efforts towards an LLVM-based intermediate representation, which could allow re-using LLVM-based verifiers.
Other semantics-based approaches rely on the structural operational semantics of the source language,
such as~\cite{DBLP:conf/ppdp/AngelisFPP15},
where verification conditions are generated in the form of Horn clauses, and~\cite{DBLP:journals/scp/Vidal15},
that focusses on termination analysis; both techniques are tied to a specific analysis technique.
All the cited semantics-based approaches were mostly developed to analyse software,
but it might be possible to adapt them to domain-specific contexts.

Further efforts in the direction of semantics-based analysis go on to decouple for good
the specification language from the actual analysis technique by means of mechanised encodings
from domain-specific languages to general-purpose imperative languages~\cite{DBLP:journals/tosem/StefanoNI22}.
This methodology has been used to analyse different emergent properties of a selection of classes of systems,
such as foraging ants~\cite{DBLP:conf/cmsb/NicolaSIV23}, flocking birds~\cite{DBLP:conf/isola/NicolaSIV22}, and emergent synchronisation in applauding audiences~\cite{DBLP:conf/isola/StefanoI24}.
While this technique allows to leverage a comprehensive range of mature verification techniques 
developed for software verification, a certain loss of structure is inevitably introduced 
during the translation from the domain-specific input language.
The structure is further manipulated when further converting the translated specifications 
into a suitable input for a general-purpose decision procedure.
Preliminary attempts to extract part of the structure upfront and use it later on in 
the verification flow to override the standard heuristics of the decision procedures
have been carried out with SAT-based analysis,
extending the range of simulation up to a few hundreds of epochs~\cite{DBLP:conf/isola/NicolaSIV22}.
This suggests that tailored heuristics for general-purpose decision procedures
can considerably improve analysis efficiency. 
%\tocheck{
%++++parallel analysis.
%++++efforts on the whole workflow, cannot do in isolation.
%++++tradeoff between separation of concerns between specification and decision procedure and
%tighter integration for tailored approaches aimed at efficiency gains.
%}

%% added by dragisa, reviewing 
\subsection{Towards a research program for rigorous market core engineering}

\textcolor{blue}{The discussion above points not only to a methodological difficulty, but also to a concrete program for formal market core engineering. The objective is to develop specification and analysis frameworks in which the interaction rules of trading venues are explicit, machine-tractable, and reusable across related mechanisms. This direction is already grounded in ongoing work on concrete cases, including Reaction Systems formalizations of CLOB, FBA, refined FBA, and related experimental models~\cite{hhmst24,hhmst26}, together with actor-based executable prototypes~\cite{hhmst24a}, based on join patterns~\cite{fg02}. A natural next step is to organise such efforts into a more systematic workflow: formal specification of market mechanisms, identification of mechanism level correctness properties, structure-preserving interfaces to verification technology, and executable models for simulation-based study of strategic and emergent behaviour. The point is not to claim a single definitive formalism, but to move toward a disciplined methodology in which market mechanisms can be specified, compared, analysed, and refined with a level of rigour closer to that expected for other critical computational infrastructures.}

% JOIN calculus ref, to insert
% \bibitem{Haller2024} P. Haller, A. Hussein, H. Melgratti, A. Scalas, and E. Tuosto. \emph{Fair Join Pattern Matching for Actors}. In \emph{38th European Conference on Object-Oriented Programming (ECOOP 2024)}, LIPIcs 313, 17:1--17:28, 2024.

\paragraph{Benchmark mechanisms.} \textcolor{blue}{To the best of our knowledge, a plausible blueprint for such a discipline begins with a small number of canonical market cores, treated as benchmark mechanisms for formalization and comparison. 
} 

\paragraph{Common correctness criteria.} \textcolor{blue}{The next step is to identify a corresponding catalogue of mechanism-level properties -- such as admissibility of interactions, priority preservation, persistence of unmatched orders, and well-defined stable states -- that can serve as common correctness criteria across models.}

\paragraph{Complementary semantic views.} \textcolor{blue}{On this basis, one can then relate complementary semantic views of the same mechanism, for instance resource-based, process-based, etc., so that formal reasoning and simulation are not disconnected activities but different access points to the same object of study. These formal layers can in turn support the study of strategic and emergent behaviour by making explicit which phenomena follow from the design of the core itself and which arise only from the interaction of adaptive agents with that core. In this sense, the proposed discipline is not a finished framework, but rather a structured research path toward reusable methods for the formal analysis of market mechanisms.}

%\textcolor{red}{Not sure where to add the reference: O. Inverso, E. Tuosto, and D. {\v{Z}}uni{\'c}.
%\emph{Fundamental Market Design as a Layer of AI-Agent Alignment}.
%Accepted at the EC'26 Workshop on Incentive-Based AI Alignment,
%co-located with the 27th ACM Conference on Economics and Computation,
%Rome, Italy, 2026.}

%%% Local Variables:
%%% mode: LaTeX
%%% TeX-master: "2026-ISoLA-Economics"
%%% End:

%!TEX root = 2026-ISoLA-Economics.tex

%--------- --------- --------- --------- --------- --------- --------- --------- --------- --------- 
\section{Concluding remarks}\label{sec:conc}
%--------- --------- --------- --------- --------- --------- --------- --------- --------- --------- 

So-called complex or collective systems are specific classes of concurrent systems
of particular interest in many disciplines and areas of research well beyond computer science.
Electronic financial markets yield a distinctive class of collective systems where concurrency
plays a crucial role.

Different peculiar elements of complexity make automated reasoning on these systems
particularly challenging. We have focused on the case of electronic financial markets to elaborate on 
the most challenging aspects.

Despite the advancements over the last few decades
in terms of formal methods, languages, techniques, and tools
for automated specification and analysis,
decision procedures, and hardware gains,
profitable domain-specific applications
seem to hardly reach beyond relatively simple illustrative examples and proofs of concept.
Shortage of state-of-the-art technology for computer-aided analysis persists,
with simulation remaining the most frequently used option, even though by
itself it provides limited support for investigating sophisticated
phenomena such as emergent behaviour.

We have argued that integrated frameworks for specification and analysis 
targeted at domain experts (not technical experts) may fill in this gap.
In particular, domain-specific languages may encourage usage,
but require considerable interdisciplinary effort for proper development. 

On the one hand,
the separation of concerns between the specification language and the actual analysis technique
is essential to re-use existing mature tools
for mainstream programming languages and general-purpose decision procedures
(e.g. SAT or SMT) that have made remarkable efficiency gains over the last few decades.
On the other hand,
further efforts are likely needed on a case-by-case basis
to tailor the different elements in the verification flow
(i.e. not only the specification language, but also the actual verification tool, and
 in some cases even the heuristics of the decision procedure) for successful scalable applications.

% All these efforts are strongly motivated by several potential advantages. 
% Effective technology for automated reasoning can allow deep investigations
% and facilitate potential insights for specific domains, with 
% potentially large interdisciplinary gains.
% %
% In fact, a recent direction proposed in~\cite{BBTZ2026} explores
% the benefits of concurrency in the formalisation of trading mechanisms
% of electronic market models, such as the one of FBA.
% %
% At the same time, new scenarios will certainly challenge
% state-of-the-art solutions and push the development forward,
% stimulating technical gains.

%
All these efforts are strongly motivated by several potential advantages.
Effective technology for automated reasoning can allow deep investigations
and facilitate potential insights for specific domains, with
potentially large interdisciplinary gains.
In fact, a recent direction proposed in~\cite{BBTZ2026} explores
the benefits of concurrency and parallelism in the formalisation of trading mechanisms
of electronic market models, such as the one of FBA.
At the same time, new scenarios will certainly challenge
state-of-the-art solutions and push the development forward,
stimulating technical gains.

% Added, Dragisa, after the reviews
\textcolor{blue}{More broadly, the challenge raised by emergent behaviour in financial markets is also to find a workable methodological path for reasoning about the mechanisms from which such complexity arises. We have argued that this requires formal descriptions of market cores, careful separation of concerns between specification, verification, and simulation, and better integration between domain-specific modelling and reusable formal-analysis technology. The aim is not to force all questions into a single formalism, but to move toward a disciplined methodology in which market mechanisms can be specified, compared, analysed, and refined with greater rigour.}

\textcolor{blue}{The discussion also points toward a more concrete research programme. This direction is already informed by ongoing formal and executable work on canonical market cores, including our own efforts. A natural next step is to bring these strands together into a more reusable workflow: benchmark mechanisms, common correctness criteria, structure-preserving links to verification technology, and executable models for the study of strategic and emergent behaviour. In this sense, the contribution is not only to highlight a difficult problem, but also to outline a plausible path toward a discipline of rigorous market-core engineering.}

%%% Local Variables:
%%% mode: LaTeX
%%% TeX-master: "2026-ISoLA-Economics"
%%% End:

\bibliographystyle{plain}
\bibliography{2026-ISoLA-Economics}

\end{document}